# SmartMask – Developing an automated self-care system

Ruchita Bhadre
*Dept. of Instrumentation and Control*
*College of Engineering, Pune*
Pune, India
ruchitabhadre@gmail.com

Tejas Ranka
*Dept. of Instrumentation and Control*
*College of Engineering, Pune*
Pune, India
tejasranka2807@gmail.com

Prathamesh Yeole
*Dept. of Instrumentation and Control*
*College of Engineering, Pune*
Pune, India
prathamesh.yeole28@gmail.com

Dr.Rohini Prashant Mudhalwadkar
*Dept. of Instrumentation and Control*
*College of Engineering, Pune*
Pune, India
rpm.instru@coep.ac.in

*Abstract*—COVID-19 has changed our world and has filled people with fear and anxiety. Everyone has a fear of coming in contact with people having the Coronavirus. In Spite of releasing full lockdowns, there is still a pressing need to maintain social distancing in the short- to medium-term to control the spread of coronavirus. Due to lack of self discipline or obviously pulling down the mask to get some fresh air, might pose a threat when you come near a person showing COVID symptoms. Abiding to WHO guidelines to avoid touching the mask while wearing it, we propose a wearable device for no contact pulling up of mask on face and additionally to implement social distancing with sensors mounted on the device. The SmartMask will detect if we are in the vicinity of any other person and will pull itself up. With sensors for detecting the closeness of objects around you and prompting you to take a proper action or pull the mask automatically. Along with the automated mask we will incorporate a temperature sensor to check vitals of an individual at all times and give an alert to the peers around him. This will ensure social distancing and help in avoiding spread of the virus.

## I. Introduction

Amid fears of a second peak of coronavirus, many people are scared to go out of the house even after the lockdown has been lifted. Along with this according to WHO, the primary route of transmission of SARS-CoV-2 is through facial organs. This evidently puts more pressure on protective measures to maintain social distancing and possible contraction of the virus. Since this COVID-19 pandemic has come into picture recently only, and since then many technological advances are coming in the limelight. We also did some market research about technological advances in battling and preventing spread of the coronavirus. These are few techniques which motivated us and drove us to innovate something which is simple yet very effective. As in keeping with Health Experts, Social Distancing may be an powerful degree to curtail the unfold of COVID19.Keeping far of as a minimum one meter from different humans lessens the probabilities of having inflamed with the virus. It is a prevention and manipulate intervention applied to lower touch and to sluggish down the price and quantity of disorder transmission in a community.

## II. Methodology

Due to COVID pandemic, in everyday life we have to take many precautions like maintaining social distance and wearing masks in public places and avoid touch to the facial organ. Due to lack of self discipline or obliviously pulling down the mask to get some fresh air, might pose a threat when you come near a person showing COVID symptoms. Abiding to WHO guidelines to avoid touching the mask while wearing it, we propose a wearable device "SmartMask". This device will open or close the mask as per the instruction given by the user. The opening and closing of the mask will happen with absolutely no touch with any facial organ. In addition to this device will also play an important role to maintain the social distance. When a person will come in the radius of 1 meter, the device will alert the user to maintain the proper distance.

## III. Hardware and Software

### A. PIR Sensor

A passive infrared sensor is a digital sensor that measures infrared mild radiating from items in its subject of view. They are most customarily utilized in PIR-primarily based totally movement detectors. PIR sensors are typically utilized in protection alarms and automated lighting fixtures applications. The Fresnel lens is used to look that the 2 slots of the PIR can see out beyond a few distance. When the sensor is inactive, then the 2 slots experience the equal quantity of IR. The ambient quantity radiates from the outdoors, partitions or room, etc. When a human frame or any animal passes by, it intercepts the primary slot of the PIR sensor. This reasons a high-quality differential alternate among the 2 bisects. But whilst the frame leaves the sensing area, the sensor generates a poor differential alternate among the 2 bisects.

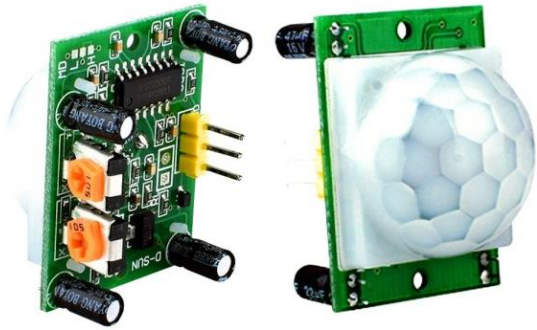

## B. IR Sensor

An infrared sensor is an electronic device that emits the light in order to sense some aspects of the surroundings. An IR sensor can measure the heat of an object as well as detects the motion. This sensor is used to detect motion of objects in the range. In this application it is used to detect the signal given by the user for opening and closing the mask.

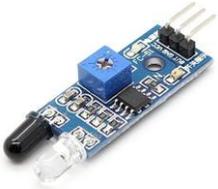

## C. Servo Motor: Tower Pro SG-90

A servo motor is a type of motor that can rotate with great precision based on the principle of Servo Mechanism. Normally this type of motor consists of a control circuit that provides feedback on the current position of the motor shaft, this feedback allows the servo motors to rotate with great precision. If you want to rotate an object at some specific angles or distance, then you use a servo motor.

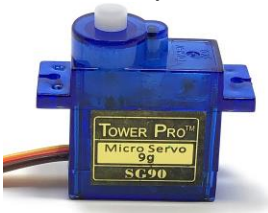

## D. Microcontroller : ESP8266 (Node-MCU)

NodeMCU is an Open-source, Interactive, Programmable, Low cost, Simple, Smart, WI-FI enabled microcontroller. It includes firmware which runs on the ESP8266 Wi- Fi SoC from Espressif Systems, and hardware which is based on the

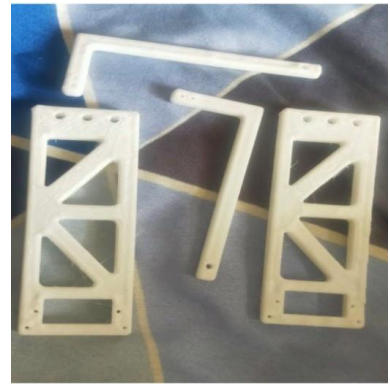

Fig. 1. 3D printed components ESP12 module.

This type of microcontroller is mainly used

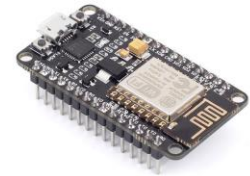

for IOT application.

## E. Power Supply

The SmartMask is powered by a Li-ION rechargeable battery. This battery is selected because :

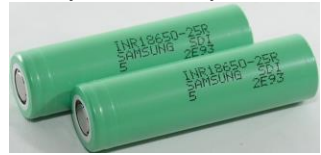

- It has a long running time.
- It requires low maintenance.
- It has an internal protection circuit.

## F. 3D Printing Model

The structure of the SmartMask is manufactured with the help of 3D printing. 3D printing model is selected because:
- Weight of the 3D components is much less.
- Servo motors can easily open and close masks using 3D printed components.
- 3D printed components provide rigid structure with good strength. See fig.1.

## G. Creo 7.0

Creo is a suite of Computer-aided design apps supporting product design. Creo delivers the most scalable range of 3D CAD product development packages and tools in today's market. Creo 7.0 has breakthrough innovations in the areas of generative design, real-time simulation, multibody design. In

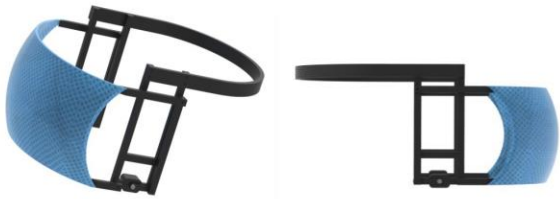

Fig. 2. Design of prototype using Creo 7.0

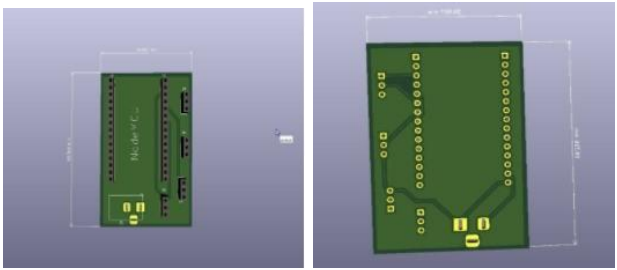

Fig. 3. PCB design

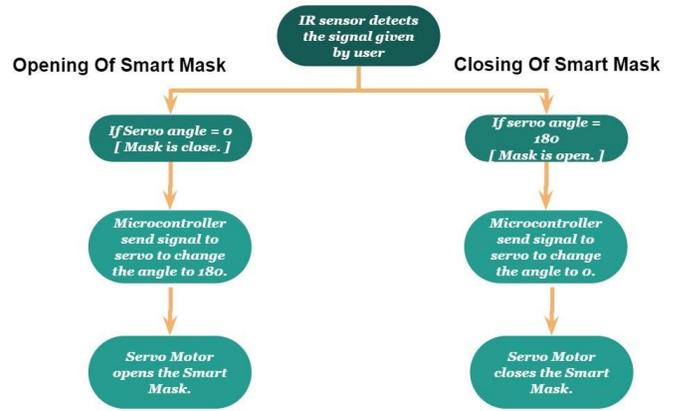

Fig. 4. Flowchart of No touch operating system

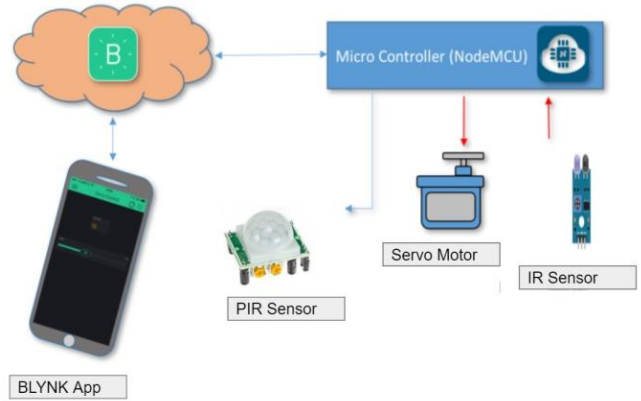

### H. KiCad EDA

KiCad EDA is a cross platform and open source electronics design automation suite. This software is mostly used for the purpose of schematic capture, PCB Design and 3D viewer of PCB. In this application we used this software for the design of PCB which includes systematic position and mounting of electronic components to avoid future problems like loose connection. The design is shown in fig.3.

this application we used this software for the 3D design of the mask. With the help of this software we manufactured an innovative design of the SmartMask.

### I. BLYNK

BLYNK is a Platform with iOS and Android apps to control any microcontroller over the internet. In this application,

- Opening and closing of masks is done through this app.
- Proper instruction and alarm is given to the user through this app.

## IV. WORKING PRINCIPLE

The SmartMask consists of mainly two systems.

### A. No touch operating system

This system mainly deals with touchless opening and closing of masks. This system (shown in fig.4) consists of NodeMCU, Servo Motor and IR sensor. We can use this system in 2 ways:

*1) By physical signal given by user:* When the user wants to open the mask, the user will move the hand in front of the IR sensor. The IR sensor will detect the motion of the hand. The IR sensor will generate the signal and this signal is received by a microcontroller. And depending upon the previous angle of servo motor, the microcontroller will give command to the servo motor to change its angle to desired angle. In this way touchless opening and closing of the mask is done.

Fig. 5. System representation of the SmartMask

*2) By BLYNK application:* When the user wants to open the mask, the user will use the BLYNK app to give the command to the microcontroller for operation of the mask. User will open the BLYNK app and the slider present in the app will give facility to the user to rotate the mask through any angle as per his convenience.

### B. Social Distancing System

This system mainly deals with giving instruction and alarm to the user when proper precaution is not taken by the user. This system consists of a PIR sensor, microcontroller and buzzer. When a person comes in the range of 1 meter of the user PIR sensor will detect the motion of the person and give instruction to the user for proper action like keep safe distance on the mobile application BLYNK.

## V. RESULTS

This section gives the perception of the exploratory outcomes. Proposed device "SmartMask : An automated self care system" is completely developed. With the help of the proposed design of the mask we manufactured, a similar mask with the help of 3D printing and tested the feasibility of the mask.

the user to open and close the mask and also provide safety instruction in public places. This

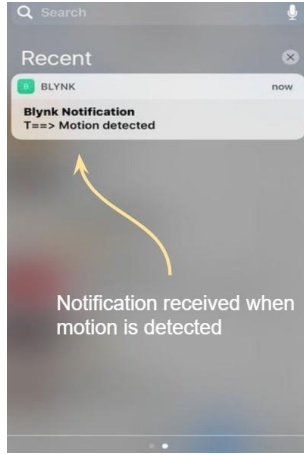

Fig. 6. BLYNK interface and social distancing notification

With the help of handmade model, we performed a trial of the proposed design and achieved the following results and specifications of the SmartMask: Achieved Specification :

1) Opening and closing of mask: From angle 0 to 180 degree
2) Detect the presence of human being : Range 1 meter
3) Low weight and innovative design for reliability as shown in fig.8 and fig.9.
4) IoT assistance through mobile app as shown in fig.6 and fig.7

## VI. CONCLUSION

The world is espousing an unprecedented technological trend for connecting billions of devices via internet and Bluetooth. The Internet of Things is a new paradigm that is evolving each day and enriching our everyday life. It also promises to drive significant changes and cause a huge impact in modern healthcare, by enabling a more customized, preventive and collaborative form of care. In this paper we presented SmartMask, an IoT-based automated self care system designed to prevent spread of COVID through facial organs. In addition to this, it helps to maintain social distance in public places. With the help of this system users can open and close the mask without having to touch their face or have any kind of contact with any facial organ. Users can open and close the mask through a simple physical action ie by waving their hand in front of the IR sensor or through the BLYNK app. This device helps the user to maintain the social distance in the public place. This device alerts the user when proper precautions are not taken and social distance is not maintained in a public place. SmartMask is a wearable device, integrated on a simple, discrete and comfortable model, offers a suitable solution to be used by any person anywhere. The BLYNK application allows device will help the community to follow the WHO guidelines and help to prevent the spread of COVID. With our SmartMask we aim to provide technological assistance to fight with this pandemic.

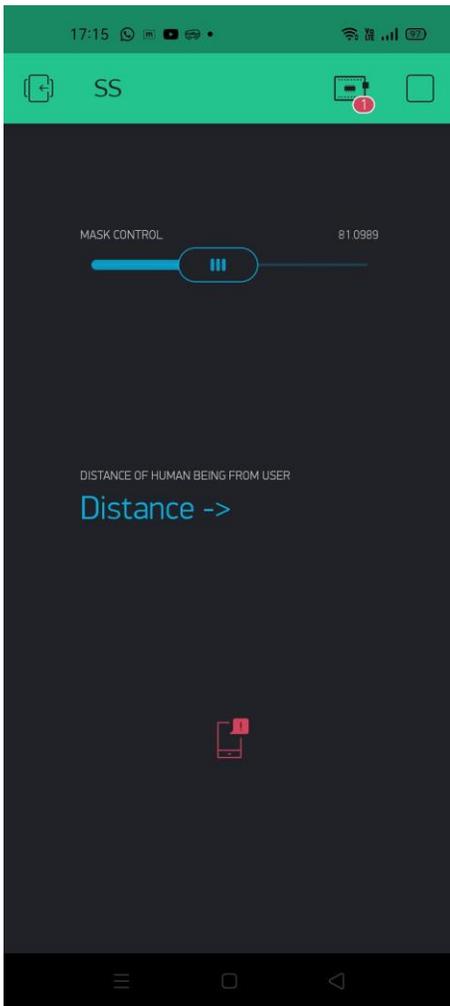

Fig. 7. IoT assistance through BLYNK app

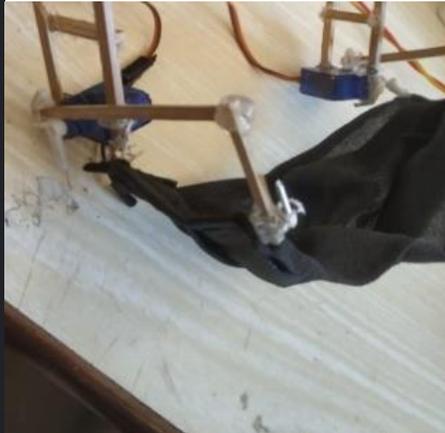

Fig. 8. Mounting of servo motor on mask

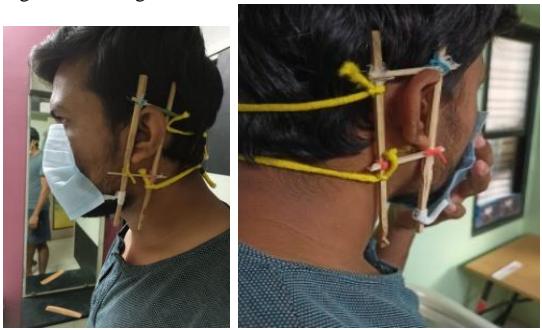

Fig. 9. Hand-made model of SmartMask

ACKNOWLEDGMENT

We would like to express our deep and sincere gratitude to our project guide Dr. Mrs. Rohini Prashant Mudhalwadkar for her constant support and valuable inputs throughout this research. We would also like to extend our heartfelt gratitude to the Head of Department of Instrumentation and Control, Dr.D.N. Sonawane.